# Fairness-Regularized DLMP-Based Bilevel Transactive Energy Mechanism in Distribution Systems

Ahmad Khaled Zarabie, *Student Member, IEEE*, Sanjoy Das, Mohammed Nazif Faqiry, *Member, IEEE*

*Abstract*—Distribution locational marginal pricing (DLMP) can adversely affect users in a grid-constrained transactive distribution system market (DSM) that are at a distance away from the substation, requiring longer paths to connect to the substation. When the grid operates closer to its physical limits in terms of line capacities and voltage deviations, these users are more likely to cause grid violations than others in the vicinity of the substation. Conversely, increased energy consumption by users near the substation can choke off the supply to those at the grid's extremities. This research describes a novel mechanism to charge users in a more equitable manner, by regularizing the distribution system operator (DSO)'s social welfare based objective function with a fairness criterion. The Jain's index of fairness is used in this context and corresponding fairness DLMP component is derived. The overall problem entails the maximization of the regularized objective within a set of linear constraints. The constraints that ensure that the grid's voltage and line power limits are not violated. Constrained optimization is carried out in an iterative manner through dual decomposition where the dual variables and unit costs are incrementally updated by the DSO using the augmented Lagrangian method (ALM). Simulations reported in this study confirm the effectiveness of the proposed approach.

## I. Nomenclature

| | |
|---|---|
| $\mathcal{N}$ | Set of nodes, lines $|\mathcal{N}| = N$ |
| $\mathcal{A}$ | Set of nodes with aggregators, $|\mathcal{A}| = A$ |
| $k, l$ | Node indices, $k, l \in \mathcal{N}$ |
| $u(k)$ | Index of $k$'s immediate upstream (parent) node |
| $d(k)$ | Set of $k$'s immediate downstream nodes |
| $r_k, x_k$ | Resistance, reactance of line $(u(k), k)$, $k \in \mathcal{N}$ |
| $p_k, q_k$ | Real, reactive power injection into node, $k \in \mathcal{A}$ |
| $c_k$ | Per unit cost of node$\pi$ $k \in \mathcal{A}$ |
| $P_k, Q_k$, | Real, reactive line power flows $k \in \mathcal{N}$ |
| $L_k^P, L_k^Q$ | Real, reactive losses $k \in \mathcal{N}$ |
| $V_k, \delta_k$ | Voltage, angle of node $k \in \mathcal{A}$ |
| $\epsilon$ | Maximum allowable pu voltage deviation |
| $u(\cdot)$ | Immediate upstream node |
| $d(\cdot)$, | Set of immediate downstream nodes |
| $\mathcal{U}(\cdot)$, | Set of all upstream nodes of given node |
| $\mathbf{D}, \mathbf{T}$ | Downstream and tree matrices |
| $\mathcal{G}_k$ | Set of agents at aggregator$k$, $k \in \mathcal{A}$, $G_k = |\mathcal{G}_k|$ |
| $i$ | Index of agent $i \in \mathcal{G}_k$ |
| $p_k^i, q_k^i$ | Real and reactive demands of $i \in \mathcal{G}_k$ |
| $g_k^i, a_k^i, b_k^i$ | Generation and utility parameters of $i \in \mathcal{G}_k$ |
| $\underline{\boldsymbol{\alpha}}, \overline{\boldsymbol{\alpha}}, \boldsymbol{\beta}, \lambda, \gamma$ | Dual variables |
| $\mathbf{c}_C, \mathbf{c}_V$ | DLMP components for congestion, voltage |
| $\mathbf{c}_{E+L}, \mathbf{c}_F$ | DLMP components for energy and loss, fairness |
| $\Omega(\cdot)$ | System level objective function |
| $\mathcal{W}_k(\cdot)$ | Social welfare of aggregator $k \in \mathcal{A}$ |
| $u_k^i(\cdot)$ | Utility of $i \in \mathcal{G}_k$ |
| $\pi_k^i(\cdot)$ | Payoff of $i \in \mathcal{G}_k$ |
| $\mathfrak{L}(\cdot)$ | Lagrange function |
| $J(\cdot)$ | Jain's index of fairness |
| $\eta$ | Increment factor per iteration |
| $C$ | Regularization weight |

In this list, only the main symbols used throughout the paper are shown. Other symbols are abstract combinations of these symbols or explained in the text. Bold lowercase and bold, italic uppercase symbols represent vectors while bold uppercase symbols represent matrices.

## II. Introduction

THE rapid proliferation of distributed energy resources (DERs) in the energy grid necessitates the need for the design of efficient transactive distribution system markets (DSMs). Pricing mechanisms that are compatible with the physical structure of the distribution grid and take operation limits into account, are being proposed in the literature [1]-[6]. Distribution locational marginal pricing (DLMP) as an effective means to establish the price of electricity in transactive DSMs has received significant research attention [2]-[6].

The latest research on DLMP-based pricing decompose locational prices to its energy, losses, voltage violation, and congestion components, [4], [5]. An inherent drawback of this method is the spatial variations in the resulting DLMP. When the grid is under stress (e.g., due to a line congestion or a node voltage hitting its operation limit), the DLMP-based pricing intrinsically charges distant nodes at higher rates than those closer to the substation. In particular, the effect of location in DLMP is substantially high if an extreme node violates the voltage constraint due to higher voltage drop.

In this paper, we propose a novel method that addresses the issue of fairness in DLMP-based DSM. We propose a fairness-regularized mechanism that can be implemented by the distribution system operator (DSO). An iterative gradient descent algorithm based on dual decomposition maximizes the global social welfare of the grid, but with spatially driven discrepancies in how the prosumers within the aggregators are charged. Incorporating fairness into the algorithms objective has a demonstrably beneficial effect on the unit costs.

*A. Related Work*

DLMP-based pricing methods and its DSM models either use DCOPF or a variant of AC optimal power flow (ACOPF) to set grid and operation limit constraints. A few papers on DCOPF-based DLMP formulations have appeared [2], [7], and [8]. References [2] and [7], propose DLMP-based methods through quadratic programming and chance-constrained mixed integer programming that use DCOPF to define line congestion and alleviate it through dynamic tariffs. The work in [8] proposes two benchmark pricing methodologies, namely DLMP and iterative DLMP (iDLMP), for a congestion free energy management by buildings providing flexible demand. Aggregators are assumed to have contracts with flexible buildings to decide their reserve and energy schedule by interacting with the DSO in a cost optimal manner in order to avoid congestion in the distribution grid. Unfortunately, due to lower x/r ratio in the distribution system, DCOPF-based DLMP has been shown to introduce significant errors [3]. Moreover, these techniques lack certain key features such as losses, voltage deviations, and reactive power pricing essential in transactive DSM.

ACOPF-based DLMP formulation to determine one or more of its constituents such as energy, loss, voltage violation, and congestion prices are investigated in [3], [4], [5]. In [3], a novel LPF method is proposed. In this paper, real and reactive energy and loss portions of the DLMP are derived. In [4], a DSM model with DLMP clearing has been proposed to manage congestion and provide voltage support. This paper uses a mixed-integer second-order-cone programming to model ACOPF, and determines binary variables such as feeder configuration status and tap locations of shunt capacitors and transformers. Similarly, in [5], authors use a trust-region based solution methodology to obtain DLMP and its constituents through a first-order approximation of the AC power flow manifold model in [5]. This paper shows better performance over [3].

What is absent from the above research is the issue of fairness in DLMP pricing mechanisms. Fairness considerations in other forms of pricing have begun to be addressed recently. Several papers, [9]-[12] make use of the Shapley value, a concept borrowed from coalitional game theory, to accomplish fairness. In [9] prosumers' fair hourly billings is achieved depending on how the DR meet system objectives. The price of anarchy, which is the deviation of the Nash equilibrium operating point from the optimal has been used to incorporate fairness in hourly billings to prosumers in [13], and as a basis for comparing two models of demand side management in terms of fairness in [14]. A method to determine fair energy costs to consumers based on their contribution to minimize overall system costs is proposed in [10]. Several fairness criteria based on emission minimization, minimize peak-to-average ratio, etc. have been proposed in [15]. In [11], a pricing during direct trade among prosumers energy is proposed using the Shapley value. A fairness-based criterion is proposed in [12] to share the cost savings in a coalition of prosumers equipped with renewable energy sources and energy storage systems.

The aim of these approaches is to reward users that consume energy during more desirable time intervals and conversely, to penalize those that demand energy during undesirable periods. We term this aspect of fairness as temporal fairness. The expectation of temporal fairness is typically to redirect the grid's operation towards more feasible operating regions. Unfortunately, to the best of the authors' knowledge, other aspects of fairness in pricing mechanisms is absent from these efforts. Energy consumption by a group of users in one area of the grid affects how the other users in the grid are priced. For instance, when consumers in a node that is positioned close to the substation transformer draw a disproportionately large amount of energy, the amounts that those further downstream can obtain from the grid is stymied. It is this relative advantage or drawback of the end users, which is based on their locations that this research proposes to mitigate. It does so by incorporating another form of fairness, which we term as spatial fairness. Spatial fairness can be best included within DLMP pricing. None of the existing papers in the available literature address the issue of spatial fairness in DLMP pricing. Additionally, most of the studies require some sort of off-the-shelf solver to compute DLMP and the shadow prices of the constraints in their optimization models [3], [5].

*B. Technical Contributions*

The key contributions of the approach proposed in this paper are summarized below.

(*i*) The proposed DLMP pricing mechanism addresses spatial fairness. This is accomplished by incorporating a regularization term in the system level objective function (addressed below). The Jain's index of fairness has been used for this purpose. Jain's index is an instance of a general class of fairness criteria that possess desirable features [16], [17]that render them particularly well-suited for user-centric resource allocation applications [18]. Furthermore, unlike its use as an evaluative tool [19] or as a system constraint [20], the proposed approach successfully includes Jain's index directly within the optimization algorithm, as well as in the DLMP pricing.

(*ii*) The proposed framework accommodates physical constraints of the grid. It uses a linearized power flow method. Similar approaches have been used elsewhere [1], [5], [21], [22]. Linearizing the power flow in this research, which is directly based on [3], not only helps in simplifying the underlying computations, but also allows the components of the DLMP to be readily available.

(*iii*) Prosumers in this framework are not required to reveal private information despite having their own DGs and individual utility functions. The only information exchange taking place between individual prosumers and the rest of the grid is limited to placing power demand bids in response to unit energy costs, i.e. DLMPS.

(*iv*) If the regularization term is neglected, the proposed method maximizes the social welfare of all prosumers in the grid, i.e. the sum of all their utilities. It should be noted that this task is accomplished at the DSO level despite its lack of access to the nonlinear utility functions of individual prosumers. The proof that the maximum social welfare is attained without regularization appears in a preliminary version of one aspect of this research [23].

(*v*) The proposed approach is a bi-level mechanism, with the DSO and the prosumers aiming to maximize different objectives. Prosumers are modeled as selfish agents that aim to maximize their individual payoffs, i.e. the difference between their utilities from consuming energy and the cost of procuring it from the grid (with negative demands indicating supply). Aggregators act as the interface between prosumers and the grid

(*vi*) The underlying optimization is based on dual decomposition. It uses the augmented Lagrangian method [24] at the DSO level for social welfare maximization, obviating the need for off-the-shelf solvers.

The rest of this paper is organized in the following manner. Section III provides details of the components of the framework adopted in this research, as well as the information flowing between them. A rigorous treatment of the DSO mechanism is postponed until Section IV next, which provides mathematical details of the linearized power flow equations, the Jain's fairness index as used here, followed by the overall formulation of the optimization problem. Section V describes the results obtained from simulation experiments. Finally, Section VI concludes this study.

## III. FRAMEWORK

The overall schematic of the bi-level mechanism is depicted in Fig. 1. At the upper level is the DSO which acts as the intermediary between the grid and the wholesale market. The DSO possesses physical information pertaining to the distribution grid and exchanges unit cost and power demand signals from each aggregator, $k \in \mathcal{A}$. Aggregators are located at some nodes of the grid, which follows a tree structured layout. Only a subset $\mathcal{A}$ of $\mathcal{N}$ contain aggregators. Each aggregator $k$ contains a set $\mathcal{G}_k$ of prosumers within a physical neighborhood. The information flow between an aggregator and its prosumers again pertain to unit costs and demands.

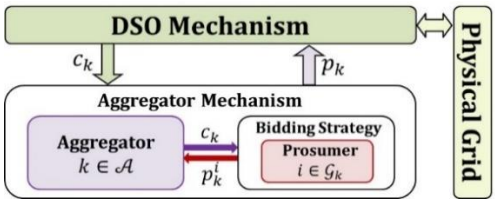

Fig. 1. Schematic of the functional components of the grid and the bi-level flow of information between them.

### A. Prosumer Agent

In a transactive DSM, contemporary retail customers are key stakeholders – prosumer agents in this framework. Each prosumer $i \in \mathcal{G}_k$ incorporates a utility function that may be construed as a measure of the satisfaction it derives from using a certain amount of energy. The utility $u_k^i : \mathbb{R}_+ \to \mathbb{R}_+$ of prosumer $i$ is a concave and strictly increasing function that is continuous and differentiable. In this research, the utilities have been modeled as logarithmic functions of the following form,

$$u_k^i(x) = a_k^i \log(b_k^i x + 1). \quad (1)$$

This is shown in Fig 2. The quantities $a_k^i$ and $b_k^i$ are prosumer specific constants, while $x$ is its load. It must be emphasized that the analytical treatment throughout this paper can handle any other utility function with the above characteristics.

Prosumers in this framework may be equipped with their own PV panels, thus capable of generating an amount of energy $g_k^i$ so that its net power demand is the difference between $x$ and generation, $g_k^i$ as,

$$p_k^i = x - g_k^i. \quad (2)$$

All parameters associated with each prosumer ($a_k^i$, $b_k^i$, $g_k^i$) as well as the choice of utility functions remain hidden from the rest of the grid.

As selfish agents, prosumers try to maximize their payoffs, which is the difference between the utility obtained from energy use and the cost of procurement. With $c_k$ being the unit cost provided to the agent, its strategy can be formulated as the following constrained optimization problem,

Maximize w.r.t. $p_k^i$:
$$\pi_k^i(p_k^i) = u_k^i(p_k^i + g_k^i) - c_k p_k^i. \quad (3a)$$

Subject to:
$$p_k^i + g_k^i \geq 0. \quad (3b)$$

Using (2), it can be shown that the demand is,

$$p_k^i = \max\left(\frac{a_k^i b_k^i - c_k}{c_k b_k^i}, 0\right) - g_k^i. \quad (4)$$

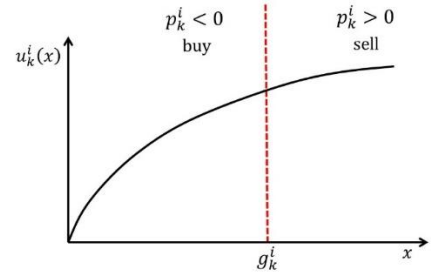

Fig. 2 Utility of a prosumer as a function of energy consumed.

### B. Aggregator

The aggregators in $\mathcal{A}$ are intermediary entities between their prosumers and the DSO, their role being primarily communicative. Each aggregator $k$ receives a unit cost signal from the latter, that it sends to the agents in $\mathcal{G}_k$. The prosumers' response is the corresponding demand $p_k^i$ as obtained from (3), or with logarithmic utilities, from (4). The social welfare at each aggregator is the sum of the utilities of all its prosumers, and given by,

$$\mathcal{W}_k(\mathbf{p}_k) = \sum_{i \in G_k} u_k^i(p_k^i + g_k^i). \quad (5)$$

In the above expression, $\mathbf{p}_k = [p_k^i]_{i \in \mathcal{G}_k}$. Neglecting the constraint in (3b) it can be seen that,

$$\frac{\partial}{\partial p_k} \mathcal{W}_k(\mathbf{p}_k) = c_k. \quad (6)$$

This shows that the aggregator $k$ responds to the DSO's unit cost $c_k$ with an aggregate energy demand $p_k = \mathbf{1}_{G_k}^T \mathbf{p}_k$ ($G_k = |\mathcal{G}_k|$) such that the slope of $\mathcal{W}_k$ is $c_k$. This information from all aggregators allows the DSO to construct the gradient $\nabla_{\mathbf{p}}[\mathcal{W}_k(p_k)]_{k \in \mathcal{A}}$ required for its optimization algorithm (for further details, one is referred to [23]).

### C. DSO

The DSO's role is in realizing the underlying optimization algorithm. It receives power demand as the vector $\mathbf{p} = [p_k]_{k \in \mathcal{A}}$ from the set $\mathcal{A}$ of aggregators and returns the DLMP $\mathbf{c} = [c_k]_{k \in \mathcal{A}}$ to the latter. Further details of the DSO are described in the next section.

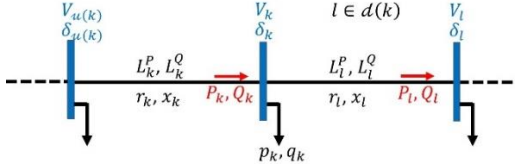

Fig. 3. Distribution system radial branch model.

## IV. MATHEMATICAL MODEL

### A. Linearized AC Power Flow

The schematic in Fig. 3 shows a segment of the radial distribution network. Each node is labeled with an index $k \in \mathcal{N}$. Line indices are identical to those of the nodes at their receiving ends. With $p_k$ and $q_k$ being the active and reactive power injected at any node $k$, the active and reactive power flowing through the line $k$ are given by the following expressions,

$$P_k = p_k + \sum_{l \in \mathcal{D}(k)} L_l^P + \sum_{l \in \mathcal{D}(k) \cap \mathcal{A}} p_l, \quad (7a)$$

$$Q_k = q_k + \sum_{l \in \mathcal{D}(k)} L_l^Q + \sum_{l \in \mathcal{D}(k) \cap \mathcal{A}} q_l. \quad (7b)$$

Here, $\mathcal{D}(k)$ is the set of all nodes that are downstream of node $k$. The quantities $L_k^P$ and $L_k^Q$ are the active and reactive line losses, that are computed as follows,

$$L_k^P = r_k \frac{P_k^2 + Q_k^2}{V_k^2}, \quad (8a)$$

$$L_k^Q = x_k \frac{P_k^2 + Q_k^2}{V_k^2}. \quad (8b)$$

In (8), $V_k$ is the voltage at node $k$ while $r_k$ and $x_k$ are the corresponding line resistance and reactance. Letting $\boldsymbol{P} = [P_k]_{k \in \mathcal{A}}$ and $\boldsymbol{Q} = [Q_k]_{k \in \mathcal{A}}$, it follows from (6) that,

$$\boldsymbol{P} = (\mathbf{I} + \mathbf{T})\mathbf{p} + \mathbf{T} L^P, \quad (9a)$$
$$\boldsymbol{Q} = (\mathbf{I} + \mathbf{T})\mathbf{q} + \mathbf{T} L^Q. \quad (9b)$$

In the above, the vector $\mathbf{p} = [p_k]_{k \in \mathcal{A}}$ may be regarded as an $|\mathcal{N}| \times 1$ vector with zeros occupying every place $k$ where $k \notin \mathcal{A}$, $\mathbf{I}$ is the $N \times N$ identity matrix. The $N \times N$ tree matrix $\mathbf{T}$ therein is defined as,

$$[\mathbf{T}]_{k, l \in \mathcal{N}} = \begin{cases} 1, & l \in \mathcal{D}(k), \\ 0, & \text{otherwise}. \end{cases}$$

The real and reactive powers at the sending end of line $k$ are given by the following expressions,

$$P_k + L_k^P = \frac{r_k V_{u(k)}(V_{u(k)} - V_k \cos(\delta_{u(k)} - \delta_k))}{r_k^2 + x_k^2}$$
$$+ \frac{V_k V_{u(k)} \sin(\delta_{u(k)} - \delta_k)}{r_k^2 + x_k^2}, \quad (10a)$$

$$Q_k + L_k^Q = \frac{x_k V_{u(k)}(V_{u(k)} - V_k \cos(\delta_{u(k)} - \delta_k))}{r_k^2 + x_k^2}$$
$$+ \frac{V_k V_{u(k)} \sin(\delta_{u(k)} - \delta_k)}{r_k^2 + x_k^2}. \quad (10b)$$

In (10), $\delta_k$ is the voltage angle at node $k$ and $u(k)$ is the index of the node that is immediately upstream of it.

The expressions in (10) are linearized to simplify the grid constraints (see later). We adopt the linearized power flow model proposed in [3]. Assuming that $|V_{u(k)}|, |V_k| \approx 1\ p.u.$, $\delta_{u(k)} - \delta_k \approx 0$, the above equalities are approximated to yield,

$$P_k + L_k^P = b_k^r(V_{u(k)} - V_k) + b_k^x(\delta_{u(k)} - \delta_k), \quad (11a)$$
$$Q_k + L_k^Q = b_k^x(V_{u(k)} - V_k) - b_k^r(\delta_{u(k)} - \delta_k). \quad (11b)$$

Here,

$$b_k^r = \frac{r_k}{r_k^2 + x_k^2}, \quad b_k^x = \frac{x_k}{r_k^2 + x_k^2}.$$

The expressions in (11) can be represented more concisely in the following manner. With $\boldsymbol{V}$ and $\boldsymbol{\delta}$ representing the $N \times 1$ vectors of all node voltages and angles, it can be shown that,

$$\begin{bmatrix} \boldsymbol{P} \\ \boldsymbol{Q} \end{bmatrix} = \mathbf{M} \begin{bmatrix} \boldsymbol{V} \\ \boldsymbol{\delta} \end{bmatrix} - \begin{bmatrix} \boldsymbol{L}^P \\ \boldsymbol{L}^Q \end{bmatrix} + \mathbf{N}. \quad (12)$$

In (12),

$$\mathbf{M} = \begin{bmatrix} \mathbf{B}^r(\mathbf{D}^T - \mathbf{I}) & \mathbf{B}^x(\mathbf{D}^T - \mathbf{I}) \\ \mathbf{B}^x(\mathbf{D}^T - \mathbf{I}) & -\mathbf{B}^r(\mathbf{D}^T - \mathbf{I}) \end{bmatrix}, \quad (13)$$

$$\mathbf{N} = \begin{bmatrix} \mathbf{B}^r V_0 \mathbf{e} + \mathbf{B}^x \delta_0 \mathbf{e} \\ \mathbf{B}^x V_0 \mathbf{e} + \mathbf{B}^r \delta_0 \mathbf{e} \end{bmatrix}, \quad (14)$$

$$\mathbf{B}^r = \text{diag}\left(\frac{r_k}{r_k^2 + x_k^2}\right), \quad (15a)$$

$$\mathbf{B}^x = \text{diag}\left(\frac{x_k}{r_k^2 + x_k^2}\right). \quad (15b)$$

The matrix $\mathbf{D}$ called the downstream matrix, is an $N \times N$ matrix defined as,

$$[\mathbf{D}]_{kl} = \begin{cases} 1, & l \in d(k), \\ 0, & \text{otherwise}. \end{cases}$$

The substation bus is indexed $0 \notin \mathcal{N}$. Its voltage $V_0$ and angle $\delta_0$ are treated as constant quantities.

Node voltages in terms of nodes real and reactive power injections, can be obtained from (9) and (12).

$$\begin{bmatrix} \boldsymbol{V} \\ \boldsymbol{\delta} \end{bmatrix} = \mathbf{C}\left(\begin{bmatrix} \mathbf{T} + \mathbf{I} & 0 \\ 0 & \mathbf{T} + \mathbf{I} \end{bmatrix} \begin{bmatrix} \boldsymbol{L}^P \\ \boldsymbol{L}^Q \end{bmatrix} - \begin{bmatrix} \mathbf{B}^r V_0 \mathbf{e} + \mathbf{B}^x \delta_0 \mathbf{e} \\ \mathbf{B}^x V_0 \mathbf{e} + \mathbf{B}^r \delta_0 \mathbf{e} \end{bmatrix}\right)$$
$$+ \mathbf{C}\begin{bmatrix} (\mathbf{T} + \mathbf{I})\mathbf{A} & 0 \\ 0 & (\mathbf{T} + \mathbf{I})\mathbf{A} \end{bmatrix} \begin{bmatrix} \mathbf{p} \\ \mathbf{q} \end{bmatrix}. \quad (16)$$

In (16), $\mathbf{e}$ is a vector with a 1 appearing as its first entry and all others being zeroes. The matrix $\mathbf{C}$ is obtained according to,

$$\mathbf{C} = \begin{bmatrix} \mathbf{B}^r(\mathbf{D}^T - \mathbf{I}) & \mathbf{B}^x(\mathbf{D}^T - \mathbf{I}) \\ \mathbf{B}^x(\mathbf{D}^T - \mathbf{I}) & -\mathbf{B}^r(\mathbf{D}^T - \mathbf{I}) \end{bmatrix}^{-1}. \quad (17)$$

By applying Taylor's series expansion around the reference points, $\mathbf{p}_0$, $\mathbf{q}_0$, $\mathbf{L}_0^P$, and $\mathbf{L}_0^Q$, linear expressions for the losses can be obtained from (2) as shown below,

$$\mathbf{L}^P = \mathbf{L}_0^P + \mathbf{J}_P^{L\,\mathrm{T}}(\mathbf{p} - \mathbf{p}_0) + \mathbf{J}_P^{L\,\mathrm{T}}\mathrm{diag}(\boldsymbol{\theta})^{-1}(\mathbf{q} - \mathbf{q}_0), \quad (18a)$$
$$\mathbf{L}^Q = \mathbf{L}_0^Q + \mathbf{J}_Q^{L\,\mathrm{T}}(\mathbf{p} - \mathbf{p}_0) + \mathbf{J}_Q^{L\,\mathrm{T}}\mathrm{diag}(\boldsymbol{\theta})^{-1}(\mathbf{q} - \mathbf{q}_0). \quad (18b)$$

The matrices $\mathbf{J}_P^L$ and $\mathbf{J}_Q^L$ in (18) are Jacobians of the losses in (8), so that,

$$\mathbf{J}_P^L = \left[\frac{\partial L_l^P}{\partial p_k}\right]_{l \in \mathcal{N}, k \in \mathcal{A}}, \qquad \mathbf{J}_Q^L = \left[\frac{\partial L_l^Q}{\partial q_k}\right]_{l \in \mathcal{N}, k \in \mathcal{A}}.$$

### B. Jain's Fairness Index

The Jain's fairness index is given by,

$$J(\mathbf{x}) = \frac{1}{1 + \hat{c}_v^2}. \quad (19)$$

In (19), $\mathbf{x} = [x_i]_{i=1:n}$ where each $x_i$ is an amount of resource allocated to any user $i$ in a set of $n$ users and $\hat{c}_v$ is the (biased) coefficient of variation of the resources. The index $J$ lies in the interval $[0,1]$ with higher values indicating more fairness. This index can be applied to the present context in a variety of ways. The simplest manner to implement fairness would be to replace each $x_i$ above, with an aggregator power $p_k$. In this manner, the fairest allocation would be when all aggregators receive an equal amounts of power. Unfortunately, this over-simplistic version of fairness does not account for the difference in the numbers of prosumers in each aggregator. However, one important insight can be gained from the above. Supplying resources to $m$ out of the $n$ users ($m < n$) and allocating $x_i = 0$ to the remaining $n - m$ would lead to a fractional Jain's index of $J = m/n$. This is one of the reasons why the aggregators that supply energy are disregarded in computing the fairness.

In this research, Jain's fairness is determined according to the expression below,

$$J(\mathbf{n} \circ \mathbf{p}) = \frac{1}{\|\mathbf{z}\|_1} \frac{(\mathbf{n}^\mathrm{T}\mathbf{p})^2}{\|\mathbf{n} \circ \mathbf{p}\|^2}. \quad (20)$$

In (20), $\mathbf{z}$ is a logical vector of 0 and 1, given by,

$$\mathbf{z} = [p_k > 0]_{k \in \mathcal{A}}. \quad (21)$$

Hence, $\|\mathbf{z}\|_1$ is the number of aggregators that receive energy from the grid. Those that supply energy ($p_k < 0$) are set aside from fairness considerations. The vector $\mathbf{n}$ is obtained as the following Hadamard product,

$$\mathbf{n} = \mathbf{c}^{\circ -1} \circ [G_k]_{k \in \mathcal{A}}^{\circ -1} \circ \mathbf{z}. \quad (22)$$

In the absence of any information regarding the size or electricity needs of the household associated with each prosumer, it is assumed that all prosumers have identical demands. Hence $G_k = |\mathcal{G}_k|$ is used for in order to allocate power to the aggregators that are roughly in proportion to their numbers of prosumers.

The formulation provided in (21) and (22) applies proportional fairness to the remaining aggregators, relies on proportional fairness, a game theoretic concept where agents in an aggregator choose to pay more for energy are allocated greater amounts of power. Its effectiveness has been demonstrated in [24].

### C. DSO Mechanism

The objective of the DSO is two-fold. It primarily attempts to maximize the social welfare $\mathbf{1}_A^\mathrm{T}[\mathcal{W}_k(p_k)]_{k \in \mathcal{A}}$ of all prosumers present in the grid. Next, it tries to price the agents as fairly as possible. Hence, the DSO's objective consists of a social welfare term, and an optional regularization term which is the Jain's fairness index in (20), weighted appropriately. Accordingly, the DSO mechanism can be formulated as the following constrained optimization problem (COP),

Maximize w.r.t. $\mathbf{p}$,

$$\Omega(\mathbf{p}) = \mathbf{1}_A^\mathrm{T}[\mathcal{W}_k(p_k)]_{k \in \mathcal{A}} + \frac{C}{2}J(\mathbf{n} \circ \mathbf{p}). \quad (23a)$$

Subject to,

$$-\mathbf{C}^V\mathbf{p} + \mathbf{c}_l^V \leq \mathbf{0}, \quad (23b)$$
$$\mathbf{C}^V\mathbf{p} + \mathbf{c}_u^V \leq \mathbf{0}, \quad (23c)$$
$$\mathbf{C}^S\mathbf{p} + \mathbf{c}_0^S \leq \mathbf{0}, \quad (23d)$$
$$\mathbf{c}^{P_0\,\mathrm{T}}\mathbf{p} + c_0^{P_0} - P_0 = 0, \quad (23e)$$
$$-\mathbf{c}^\mathrm{T}\mathbf{p} + c_0 P_0 \leq 0. \quad (23f)$$

The linear equality and inequality constraints (23b) – (23f) are obtained directly from the linearization described in Section IV.A, with appropriate rearrangements. As this research used real power to determine unit costs, they have been expressed compactly in terms of the real power which is the primal variable $\mathbf{p} = [p_k]_{k \in \mathcal{A}}$. The inequalities in (23b) and (23c) restrict the node voltage deviations to lie within $V_0 \pm \epsilon$, where $V_0$ is the voltage substation voltage. Likewise, real, and reactive line flow limits are imposed by means of (23d). The equality appearing in (23e) is the power balance constraint that ensures that $P_0$, the total power supplied to the DSO by the wholesale market equals the sum of the total power demands of the grid's prosumers and the power losses. Lastly, (23f) is the weak budget balance constraint that restricts the feasible region to one where the DSO payment at a unit cost of $c_0$ does not exceed the total monetary amount that it receives from the aggregators.

With $\underline{\boldsymbol{\alpha}}, \overline{\boldsymbol{\alpha}}, \boldsymbol{\beta}, \lambda, \gamma$ being dual variables and $\eta$, an increment factor of the augmented Lagrange function, the DSO's COP is expressed as follows,

$$\mathfrak{L}_a(\mathbf{p}, \underline{\boldsymbol{\alpha}}, \overline{\boldsymbol{\alpha}}, \boldsymbol{\beta}, \lambda, \gamma) = \sum_{k \in \mathcal{A}} \mathcal{W}_k(p_k) + \frac{C}{2}J(\mathbf{n} \circ \mathbf{p})$$
$$-\underline{\boldsymbol{\alpha}}^\mathrm{T}(-\mathbf{C}^V\mathbf{p} + \mathbf{c}_l^V) - \frac{\eta}{2}(-\mathbf{C}^V\mathbf{p} + \mathbf{c}_l^V)^\mathrm{T}(-\mathbf{C}^V\mathbf{p} + \mathbf{c}_l^V)_+$$
$$-\overline{\boldsymbol{\alpha}}^\mathrm{T}(\mathbf{C}^V\mathbf{p} + \mathbf{c}_u^V) - \frac{\eta}{2}(\mathbf{C}^V\mathbf{p} + \mathbf{c}_u^V)^\mathrm{T}$$
$$-\boldsymbol{\beta}^\mathrm{T}(\mathbf{C}^S\mathbf{p} + \mathbf{c}_0^S) - \frac{\eta}{2}(\mathbf{C}^S\mathbf{p} + \mathbf{c}_0^S)^\mathrm{T}(\mathbf{C}^S\mathbf{p} + \mathbf{c}_0^S)_+$$
$$-\lambda\left(\mathbf{C}^{P_0\,\mathrm{T}}\mathbf{p} + c_0^{P_0} - P_0\right) - \frac{\eta}{2}\left(\mathbf{C}^{P_0\,\mathrm{T}}\mathbf{p} + c_0^{P_0} - P_0\right)^2$$
$$-\gamma(-\mathbf{c}^\mathrm{T}\mathbf{p} + c_0 P_0)$$
$$-\frac{\eta}{2}(-\mathbf{c}^\mathrm{T}\mathbf{p} + c_0 P_0)(-\mathbf{c}^\mathrm{T}\mathbf{p} + c_0 P_0)_+. \quad (24)$$

Equating its derivative $\nabla_{\mathbf{p}}\mathfrak{L}(\mathbf{p}, \lambda, \xi, \gamma)$ to zero, and using (6), an incremental update rule for the unit cost is obtained as,

$$\mathbf{c}^{\mathrm{new}} \leftarrow -\mathbf{C}^{V\,\mathrm{T}}\underline{\boldsymbol{\alpha}} - \eta\mathbf{C}^{V\,\mathrm{T}}(-\mathbf{C}^V\mathbf{p} + \mathbf{c}_l^V)_+$$
$$+\mathbf{C}^{V\,\mathrm{T}}\overline{\boldsymbol{\alpha}} + \eta\mathbf{C}^{V\,\mathrm{T}}(\mathbf{C}^V\mathbf{p} + \mathbf{c}_u^V)_+$$
$$+\mathbf{C}^{S\,\mathrm{T}}\boldsymbol{\beta} + \eta\mathbf{C}^{S\,\mathrm{T}}(\mathbf{C}^S\mathbf{p} + \mathbf{c}_0^S)_+$$
$$+\lambda\mathbf{C}^{P_0} + \eta\left(\mathbf{C}^{P_0}\mathbf{p}^\mathrm{T} + c_0^{P_0} - P_0\right)\mathbf{C}^{P_0\,\mathrm{T}}$$

$$-\gamma\mathbf{c} - \eta\mathbf{c}(-\mathbf{c}^T\mathbf{p} + c_0 P_0)_+$$
$$-\frac{C}{2}\nabla_\mathbf{p} J(\mathbf{n} \circ \mathbf{p}). \tag{25}$$

In the above expression $\nabla_\mathbf{p} J$ is the derivative of the Jain's fairness index in (20), with its argument $\mathbf{n} \circ \mathbf{p}$ restricted to the positive orthant. It is given by,

$$\nabla_\mathbf{p} J(\mathbf{n} \circ \mathbf{p}) = \frac{2}{\|\mathbf{n} \circ \mathbf{p}\|}\mathbf{n} \circ \left(\sqrt{J(\mathbf{n} \circ \mathbf{p})}\frac{1}{\|\mathbf{z}\|_1}\right.$$
$$\left.-J(\mathbf{n} \circ \mathbf{p})\frac{\mathbf{n} \circ \mathbf{p}}{\|\mathbf{n} \circ \mathbf{p}\|}\right). \tag{26}$$

The DLMP components of the updated cost $\mathbf{c}^{\text{new}}$ in (25) can be readily obtained. Its DLMP components are the voltage component, $\mathbf{c}_V$, the congestion component, $\mathbf{c}_C$, as well as the energy and loss component $\mathbf{c}_{E+L}$. Additionally, regularization introduces a new fairness component, $\mathbf{c}_F$. These are as follows,

$$\mathbf{c}_V = -\mathbf{C}^{V^T}\underline{\alpha} - \eta\mathbf{C}^{V^T}(-\mathbf{C}^V\mathbf{p} + \mathbf{c}_l^V)_+$$
$$+\mathbf{C}^{V^T}\overline{\alpha} + \eta\mathbf{C}^{V^T}(\mathbf{C}^V\mathbf{p} + \mathbf{c}_u^V)_+, \tag{27a}$$
$$\mathbf{c}_C = \mathbf{C}^{S^T}\boldsymbol{\beta} + \eta\mathbf{C}^{S^T}(\mathbf{C}^S\mathbf{p} + \mathbf{c}_0^S)_+, \tag{27b}$$
$$\mathbf{c}_{E+L} = \lambda\mathbf{C}^{P_0} + \eta(\mathbf{C}^{P_0}\mathbf{p}^T + c_0^{P_0} - P_0)\mathbf{C}^{P_0^T}. \tag{27c}$$
$$\mathbf{c}_F = \frac{C}{2}\nabla_\mathbf{p} J(\mathbf{n} \circ \mathbf{p}) \tag{27d}$$

The unit cost, $\mathbf{c}^{\text{new}}$ is therefore the sum of its components,

$$\mathbf{c}^{\text{new}} = \mathbf{c}_V + \mathbf{c}_C + \mathbf{c}_{E+L} + \mathbf{c}_F. \tag{28}$$

The budget balance term is not provided as in the simulations described in the next section, the constraint was inactive at convergence, although the DSO's surplus was negligible.

The dual variables are updated using dual gradient descent as follows,

$$\underline{\alpha} \leftarrow [\underline{\alpha} + \eta(-\mathbf{C}^V\mathbf{p} + \mathbf{c}_l^V)]_+, \tag{29a}$$
$$\overline{\alpha} \leftarrow [\overline{\alpha} + \eta(\mathbf{C}^V\mathbf{p} + \mathbf{c}_u^V)]_+, \tag{29b}$$
$$\boldsymbol{\beta} \leftarrow [\boldsymbol{\beta} + \eta(\mathbf{C}^S\mathbf{p} + \mathbf{c}_0^S)]_+, \tag{29c}$$
$$\lambda \leftarrow \lambda + \eta(\mathbf{C}^{P_0}\mathbf{p} + c_0^{P_0} - P_0), \tag{29d}$$
$$\gamma \leftarrow \gamma + \eta(-\mathbf{c}^T\mathbf{p} + c_0 P_0). \tag{29e}$$

The algorithm terminates only when the updates to $\mathbf{p}$ are such that $\|\Delta\mathbf{p}\|_1 \ll 1$ for several consecutive iterations. The following proposition argues that termination occurs when the global maximum of $\Omega(\mathbf{p})$ is reached.

*Conjecture*: The objective $\Omega(\mathbf{p})$ has a unique global maximum.

*Rationale*: It can be shown that the $\nabla J(\mathbf{p}) \geq \mathbf{0}$ and when $\mathbf{p}$ is restricted to the positive orthant, $\nabla^2 J(\mathbf{p}) \leq \mathbf{0}$. Whence $J(\cdot)$ is concave and increasing. Under our assumption, the utility $u_k^i$ are concave and strictly increasing (see Fig. (2)), so is their sum $\mathcal{W}_k$ at each aggregator level, and the social welfare term in (23a). It follows that $\Omega(\mathbf{p})$ has a unique maximum.

## V. SIMULATION RESULTS

The model used here was implemented on a modified IEEE 37-bus system as shown in Fig. 4. There were 17 nodes with aggregators (shaded circles in Fig. 4). For clarity, the aggregators were indexed separately as $A1 - A17$. Three separate scenarios were created for this study. In Scenario-I all aggregators had $G_k = 10$ prosumers, without generation, but with $a_k^i, b_k^i$, generated randomly in each case. Scenario-II was similar to Scenario-I, except that the number of prosumers was doubled in aggregators $A3, A9, A11, A17$ ( $G_k = 20, k = 12, 25, 27, 36$ ). In Scenario-III prosumers in aggregators $A8, A10, A14$ ( $k = 23, 26, 31$ ) were equipped with PV generation. Their generations, $g_k^i$ were obtained randomly. All simulations were performed using MATLAB. The bi-level mechanism was investigated for each scenario, both without and with fairness regularization ($C = 0, C = 0.4$ in (23a)).

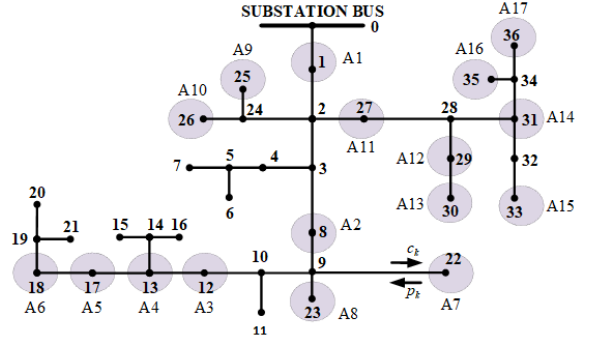

Fig. 4. IEEE 37 bus system with aggregators indexed A1 – A17.

Fig. 5 (top) shows simulation results obtained from Scenarios-I, II, and III. The vertical bars are the power allocations of the aggregators that were obtained from the simulations. Those without fairness regularization appear in blue, ($p_k$) while those with regularization are in yellow ($p_k^*$). The unit costs of the aggregators without fairness ($c_k$) and with fairness ($c_k^*$) are also provided in solid and dotted lines.

From Fig. 4 it can be seen that aggregators $A1, A9, A10$ are positioned close to the substation bus. Consequently, when not regularized for fairness the mechanism outputs costs where these aggregators are sold energy at lower rates and therefore enjoy higher power allocations. In contrast, $A5, A6, A15, A16, A17$ experience higher unit costs and lower power allocations. Fig. 5 shows how regularization helps in mitigating this adverse effect. Fairness causes aggregators to be charged in a more equitable manner.

The breakdown of the unit costs into its DLMP components is also shown in Fig. 5 (bottom). These are shown as stacked bars colored purple for the fairness component ($\mathbf{c}_F$), light blue for the unit cost due to energy usage and loss ($\mathbf{c}_{E+L}$), green for unit costs of congestion ($\mathbf{c}_C$), and yellow when the voltage limit constraints ($\mathbf{c}_V$) are active.

The DLMP components in Fig. 5 sheds further insights into the discrepancies in the unit costs. It is seen that the voltage components are very low for the three aggregators $A1, A9, A10$ located close to the substation. This is because voltage constraints are active further downstream in the grid, causing an increase in the voltage cost components in those aggregators ($A5, A6, A15, A16, A17$).

These are the aggregators that are furthest from the substation (Fig. 4). The mitigating effect of regularizing the DSO's objective is evident from the DLMP components in Fig. 5. The algorithm provides discounted unit costs to the spatially disadvantaged aggregators, which is made up by incrementing the three aggregators in the substation's neighborhood.

Fig. 6 shows results of Scenario-II with twice the agents present in aggregators $A3, A9, A11, A17$ as elsewhere, allowing one to examine the effects of congestion. The rationale behind this choice of aggregators is to the wide range of their locations vis-à-vis the substation, with $A9, A11$ being closest to it, $A3$ further away and $A17$ located at a large distance.

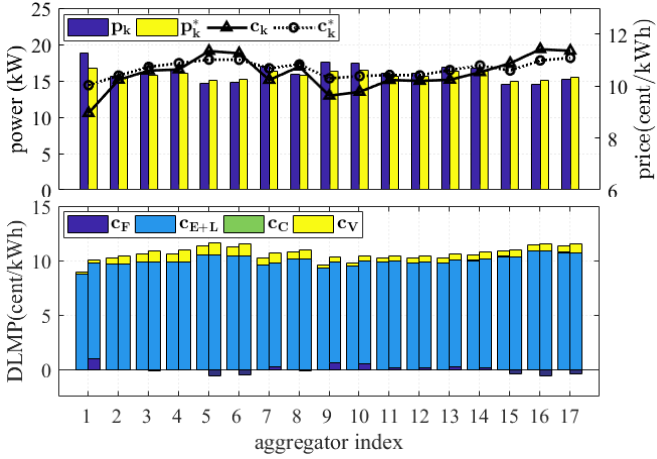

Fig. 5. Results of Scenario-I showing unit costs and allocated power to each aggregator (top) and the DLMP components (bottom).

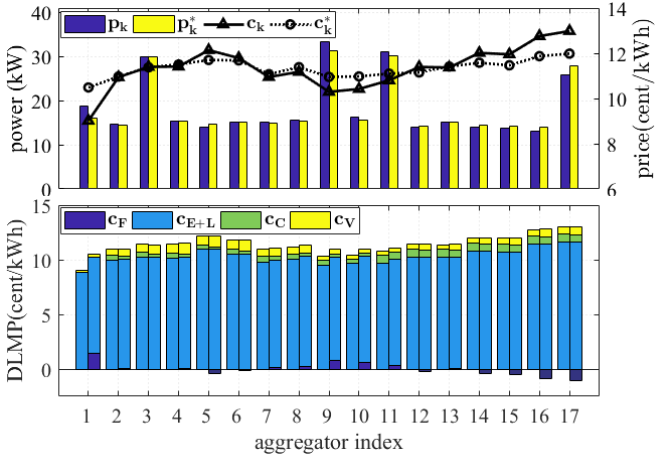

Fig. 6. Results of Scenario-II showing unit costs and allocated power to each aggregator (top) and the DLMP components (bottom).

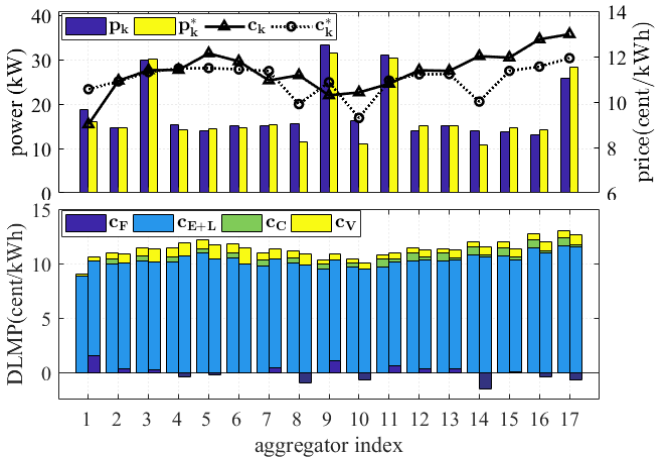

Fig. 7. Results of Scenario-II showing unit costs and allocated power to each aggregator (top) and the DLMP components (bottom).

A similar pattern as before is observed in Fig. 6, with distances having a severe impact on the unit costs. Increased loads in some aggregators causes congestion. Consequently, in the absence of regularization, aggregators yields higher unit costs for aggregators $A4, A5, A6, A15, A16, A17$ due to their distance than the others, as well as in comparison to what they were charged in Scenario-I.

The DLMP components elucidate the effect of higher congestion. Aggregators $A4, A5, A6, A15, A16, A17$ are priced at higher levels ($\mathbf{c}_C$). The significantly lower unit cost of A1 and to a lesser extent, $A9, A10$ due to their closeness to the substation, is evident. Supplementing the DSO's objective with Jain's index helps alleviate the pricing disparity. The previously advantaged aggregators see the highest unit costs due to the fairness component ($\mathbf{c}_F$). Conversely, those furthest away are able to increase their demands due to the lowered costs.

The effects of the penetration of PV generation in aggregators $A8, A10, A14$ diminishes the undesirable influence of congestion in voltage deviations throughout the distribution grid. Due to their PV generations, and resultant lower demands, the fairness component rewards them with highest drop in unit costs. This is seen in Fig. 7. Other aggregators also benefit from the introduction of PV generation in the grid in comparison the unit costs without fairness shown in Fig. 7. The congestion costs ($\mathbf{c}_C$) are uniformly lower than in the previous case. The DLMP components of the unit costs in this figure shows how regularization tries to rectify locational discrepancies.

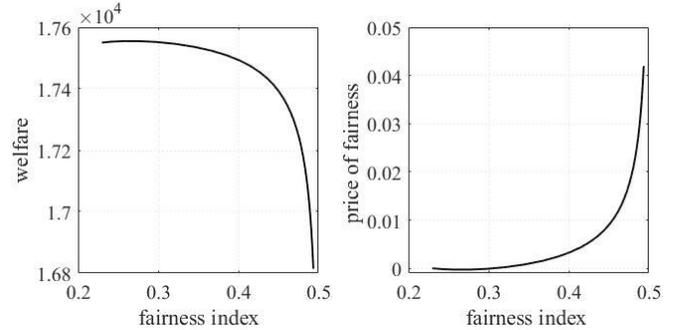

Fig. 8. Total welfare and price of fairness vs Jain's fairness index

Lastly, the tradeoff between welfare and fairness, which is very well quantified in econometric and game-theory literature, is briefly addressed. Scenario-II was simulated when the regularization weight $C$ was varied between $C = 0$ and $C = 0.5$ in increments of $0.02$. The results are shown in Fig. 8 clearly illustrating the tradeoff. Increasing fairness $J(\mathbf{n} \circ \mathbf{p})$ is associated with a simultaneous decrease in the social welfare, $\mathbf{1}_A^T [\mathcal{W}_k(p_k)]_{k \in \mathcal{A}}$, as seen in Fig. 8 (left). Fig. 8 (right) shows the same phenomenon in terms of the price of fairness – the reduction in social welfare from regularization expressed as a fraction of the maximum social welfare sans regularization,

$$1 - \frac{J(\mathbf{n} \circ \mathbf{p})}{\max_{\mathbf{p}} \mathbf{1}_A^T [\mathcal{W}_k(p_k)]_{k \in \mathcal{A}}}.$$

Fortunately, even in the extreme case ($C = 0.5$), the price of fairness is approximately equal to 0.04, i.e. an acceptable 4% reduction in the entire grid's social welfare.

## VI. Conclusions

This research entails innovations along several directions. It is privacy preserving, in that it attains the global maximum of $\Omega(\mathbf{p})$ in (23), without any knowledge about the prosumers' parameters. This is because the bidding process taking place between the prosumers and the aggregators allows the gradient $\nabla_{\mathbf{p}}\Omega(\mathbf{p})$ to be computed. The use of dual decomposition helps in dissociating the various DLMP component of the unit costs.

The application of Jain's fairness index in a distribution system and derivation of its component for DLMP is, to the best of the authors' knowledge, novel. It was shown to apportion energy to the aggregators in a more equitable manner. In addition, again to the best of the authors' knowledge, the use of Jain's index within the gradient descent algorithm is novel.

There are a few limitations of the research described here. The rate of convergence was not investigated properly. When the initialization was entirely arbitrarily, the gradient descent would start an infeasible region that was very far away from the active feasible manifold. Therefore the number of steps required before termination was of the order of $10^5$. A large speedup can be accomplished with enhancements to the basic ALM used herein or replacing it with a faster algorithm. However, the authors believe that in a real world setting with multiple time slots, using the outputs of each time slot $t$ as the initial values of $\mathbf{p}, \mathbf{c}$, even the dual variables to initialize the next, $t+1$, would yield several orders of magnitude speedups.

Linearization was used for mathematical convenience, computational simplicity, and as linear constraints guarantee unique maxima. However, the approximation error in the output must be quantified for a more thorough assessment of the approach's performance. However, it should be noted that the outputs of some simulations were compared with actual power flow. The largest errors were in the line voltages, which was very acceptable at the order of $10^{-3}$.

Finally, a more rigorous investigation into Jain's index needs to be performed, specifically its comparison with other possible methods to quantify fairness (such as the generalized curves in [16], [17], or entropy-based measures).

## VII. Acknowledgment

This work was partially supported by the National Science Foundation-CPS under Grant CNS-1544705.This work was partially supported by the National Science Foundation-CPS under Grant CNS-1544705.

## VIII. References

[1] M. N. Faqiry and S. Das, "Distributed Bilevel Energy Allocation Mechanism with Grid Constraints and Hidden User Information," *IEEE Trans. Smart Grid*, vol. 3053, no. c, 2017.

[2] S. Huang, Q. Wu, S. S. Oren, R. Li, and Z. Liu, "Distribution Locational Marginal Pricing Through Quadratic Programming for Congestion Management in Distribution Networks," *IEEE Trans. Power Syst.*, vol. 30, no. 4, pp. 2170–2178, 2015.

[3] H. Yuan, F. Li, Y. Wei, and J. Zhu, "Novel linearized power flow and linearized OPF models for active distribution networks with application in distribution LMP," *IEEE Trans. Smart Grid*, vol. 9, no. 1, pp. 438–448, 2018.

[4] L. Bai, J. Wang, C. Wang, C. Chen, and F. Li, "Distribution Locational Marginal Pricing (DLMP) for Congestion Management and Voltage Support," *IEEE Trans. Power Syst.*, vol. 33, no. 4, pp. 4061–4073, 2018.

[5] S. Hanif, K. Zhang, C. Hackl, M. Barati, H. B. Gooi, and T. Hamacher, "Decomposition and Equilibrium Achieving Distribution Locational Marginal Prices using Trust-Region Method," *IEEE Trans. Smart Grid*, vol. 3053, no. c, pp. 1–1, 2018.

[6] M. Caramanis, E. Ntakou, W. W. Hogan, A. Chakrabortty, and J. Schoene, "Co-optimization of power and reserves in dynamic T&D power markets with nondispatchable renewable generation and distributed energy resources," *Proc. IEEE*, vol. 104, no. 4, pp. 807–836, 2016.

[7] Z. Liu, Q. Wu, S. Oren, S. Huang, R. Li, and L. Cheng, "Distribution Locational Marginal Pricing for Optimal Electric Vehicle Charging through Chance Constrained Mixed-Integer Programming," *IEEE Trans. Smart Grid*, vol. 3053, no. c, pp. 1–1, 2016.

[8] S. Hanif, T. Massier, H. B. Gooi, T. Hamacher, and T. Reindl, "Cost Optimal Integration of Flexible Buildings in Congested Distribution Grids," *IEEE Trans. Power Syst.*, vol. PP, no. 99, pp. 1–14, 2016.

[9] Z. Baharlouei, M. Hashemi, H. Narimani, and H. Mohsenian-Rad, "Achieving optimality and fairness in autonomous demand response: Benchmarks and billing mechanisms," *IEEE Trans. Smart Grid*, vol. 4, no. 2, pp. 968–975, 2013.

[10] Z. Baharlouei and M. Hashemi, "Efficiency-fairness trade-off in privacy-preserving autonomous demand side management," *IEEE Transactions on Smart Grid*, vol. 5, no. 2. pp. 799–808, 2014.

[11] Z. Li, L. Chen, and G. Nan, "Small-scale Renewable Energy Source Trading: A Contract Theory Approach," *IEEE Trans. Ind. Informatics*, vol. 14, no. 4, pp. 1–1, 2017.

[12] A. Chis and V. Koivunen, "Coalitional game based cost optimization of energy portfolio in smart grid communities," *IEEE Trans. Smart Grid*, vol. 3053, no. c, pp. 1–11, 2017.

[13] P. Jacquot, P. Jacquot, O. Beaude, S. Gaubert, and N. Oudjane, "Analysis and Implementation of an Hourly Billing Mechanism for Demand Response Management," *IEEE Trans. Smart Grid*, vol. 3053, no. c, pp. 1–14, 2018.

[14] J. Ma, J. Deng, L. Song, and Z. Han, "Incentive Mechanism for Demand Side Management in Smart Grid Using Auction," *IEEE Trans. Smart Grid*, vol. 5, no. 3, pp. 1379–1388, 2014.

[15] S. K. Vuppala, S. Member, K. Padmanabh, and S. K. Bose, "Incorporating Fairness within Demand Response Programs in Smart Grid," *Innov. Smart Grid Technol. (ISGT), IEEE PES*, pp. 1–9, 2011.

[16] T. Lan, D. Kao, M. Chiang, and A. Sabharwal, "An Axiomatic Theory of Fairness in Resource Allocation," *INFOCOM (Extended)*, pp. 1–9, 2010.

[17] T. Hobfeld, L. Skorin-Kapov, P. E. Heegaard, and M. Varela, "Definition of QoE Fairness in Shared Systems," *IEEE Commun. Lett.*, vol. 21, no. 1, pp. 184–187, 2017.

[18] R. V. Prasad, E. Onur, and I. G. M. M. Niemegeers, "Fairness in Wireless Networks:Issues, Measures and Challenges," *IEEE Commun. Surv. Tutorials*, vol. 16, no. 1, pp. 5–24, 2014.

[19] F. Chiti, R. Fantacci, and B. Picano, "A Matching Theory Framework for Tasks Offloading in Fog Computing for IoT Systems," *IEEE Internet Things J.*, vol. PP, no. c, pp. 1–1, 2018.

[20] C. Guo, S. Member, M. Sheng, X. Wang, and Y. Zhang, "Throughput Maximization with Short-Term and Long-Term Jain's Index Constraints in Downlink OFDMA Systems," *IEEE Trans. Commun.*, vol. 62, no. 5, pp. 1503–1517, 2014.

[21] C. L. Chang and J. C. H. Peng, "A Decision-Making Auction Algorithm for Demand Response in Microgrids," *IEEE Trans. Smart Grid*, vol. 9, no. 4, pp. 3553–3562, 2018.

[22] M. N. Faqiry and S. Das, "Double Auction With Hidden User Information: Application to Energy Transaction in Microgrid," *IEEE Trans. Syst. Man, Cybern. Syst.*, pp. 1–14, 2018.

[23] A. K. Zarabie and S. Das, "Efficient Distributed DSO Auction with Linearized Grid Constraints," in *10th Conference on Innovative Smart Grid Technologies(ISGT)*, Washington DC, 2019, submitted.

[24] M. N. Faqiry and S. Das, "Double-Sided Energy Auction in Microgrid: Equilibrium under Price Anticipation," *IEEE Access*, vol. 4, pp. 3794–3805, 2016.